%% file: conference_101719.tex
\documentclass[10pt,conference]{IEEEtran}
\IEEEoverridecommandlockouts
\usepackage{cite}
\usepackage{amsmath,amssymb,amsfonts}
\usepackage{algorithmic}
\usepackage{graphicx}
\usepackage{textcomp}
\usepackage{xcolor}
\definecolor{darkgreen}{RGB}{0,100,0}
\usepackage{booktabs}
\usepackage{multirow}
\usepackage{xspace}
\usepackage{array}
\usepackage{fancyvrb}
\usepackage{soul}
\usepackage{xurl}
\usepackage{tcolorbox}
\usepackage{float}
\usepackage[shortlabels]{enumitem}
\usepackage{hyperref}
\usepackage{balance}
\usepackage{multicol}

\def\BibTeX{{\rm B\kern-.05em{\sc i\kern-.025em b}\kern-.08em
    T\kern-.1667em\lower.7ex\hbox{E}\kern-.125emX}}

\usepackage{xargs}                      
\usepackage[colorinlistoftodos,prependcaption,textsize=tiny]{todonotes}
\newcommandx{\unsure}[2][1=]{\todo[linecolor=red,backgroundcolor=red!25,bordercolor=red,#1]{#2}}
\newcommandx{\change}[2][1=]{\todo[linecolor=blue,backgroundcolor=blue!25,bordercolor=blue,#1]{#2}}
\newcommandx{\info}[2][1=]{\todo[linecolor=OliveGreen,backgroundcolor=OliveGreen!25,bordercolor=OliveGreen,#1]{#2}}
\newcommandx{\improvement}[2][1=]{\todo[linecolor=Plum,backgroundcolor=Plum!25,bordercolor=Plum,#1]{#2}}
\newcommandx{\thiswillnotshow}[2][1=]{\todo[disable,#1]{#2}}
\newcommand{\basicalert}[2]{\fbox{\bfseries\sffamily\scriptsize\color{blue} #1}{\sf\small$\blacktriangleright$\textit{\color{red} #2}$\blacktriangleleft$}}

\newcommand{\lx}[1]{\basicalert{From JLX}{#1}}

\begin{document}

\title{Adapting Knowledge Prompt Tuning for Enhanced Automated Program Repair}


\author{\IEEEauthorblockN{Xuemeng Cai}
\IEEEauthorblockA{\textit{Singapore Management University} \\
Singapore \\
xuemengcai@smu.edu.sg}
\and
\IEEEauthorblockN{Lingxiao Jiang}
\IEEEauthorblockA{\textit{Singapore Management University} \\
Singapore \\
lxjiang@smu.edu.sg}
}

\maketitle

\begin{abstract}
\input{abstract.tex}

\end{abstract}
    

\section{Introduction}
\label{sec:intro}
\input{introduction}
\section{Background and Our Adaptation}
\label{sec:background}
\input{background}

\section{Our Approach}
\label{sec:ourapp}
\input{ourapp}

\section{Experimental Design}
\label{sec:exprdesign}
\input{experimental}

\section{Experimental Results}
\label{sec:results}
\input{result}

\section{Discussion}
\label{sec:discuss}
\input{disscussion}

\section{Related Work}
\input{relatedwork}
\section{Conclusion}
\label{sec:conclude}
\input{conclusion}

\section*{Acknowledgments}
\input{acks}

\onecolumn
\begin{multicols}{2}
\bibliographystyle{IEEEtran}
\bibliography{ref}
\end{multicols}


\end{document}

%% file: abstract.tex
Automated Program Repair (APR) aims to enhance software reliability by automatically generating bug-fixing patches. 
Recent work has improved the state-of-the-art of APR by fine-tuning pre-trained large language models (LLMs), such as CodeT5, for APR. However, the effectiveness of fine-tuning becomes weakened in data scarcity scenarios, and data scarcity can be a common issue in practice, limiting fine-tuning performance. 
 To alleviate this limitation, this paper adapts prompt tuning for enhanced APR and conducts a comprehensive study to evaluate its effectiveness in data scarcity scenarios, using three LLMs of different sizes and six diverse datasets across four programming languages. 
Prompt tuning rewrites the input to a model by adding extra prompt tokens and tunes
both the model and the prompts
on a small dataset. These tokens provide task-specific knowledge that can improve the model for APR, which is especially critical in data scarcity scenarios. 
Moreover, domain knowledge has proven crucial in many code intelligence tasks, but existing studies fail to leverage domain knowledge during the prompt tuning for APR. To close this gap, we introduce knowledge prompt tuning, an approach that adapts prompt tuning with six distinct types of code- or bug-related domain knowledge for APR. 
Our work, to the best of our knowledge, is the first to adapt and evaluate prompt tuning and the effectiveness of code- or bug-related domain knowledge for APR, particularly under data scarcity settings.
Our evaluation results demonstrate that prompt tuning with knowledge
generally outperforms fine-tuning under various experimental settings, achieving an average improvement of 87.33\% over fine-tuning in data scarcity scenarios.

%% file: introduction.tex
Automated Program Repair (APR) aims to aid software developers in automating bug fixes. 
Over recent years, developers have constructed Large Pre-trained Language Models (LLMs), including models like CodeT5~\cite{wang-etal-2021-codet5} and CodeBERT~\cite{codebert}, which demonstrate promising performance across a wide range of code intelligence tasks.
Recent research~\cite{era} has investigated the effectiveness of applying LLMs to APR tasks by evaluating nine state-of-the-art models across five popular APR benchmarks, and their results demonstrate that LLMs can achieve comparable or superior performance compared to the current traditional~\cite{temp_soat} and DL-based~\cite{cure, recoder} state-of-the-art APR tools, highlighting the tremendous potential of leveraging LLMs in APR.

However, a gap remains between vanilla pre-trained models and fine-tuned models in downstream code tasks. Recently, fine-tuning LLMs, such as CodeT5~\cite{vulrepair}, CodeBERT~\cite{codebertfine} and GPT series, has become a common practice within the APR community and greatly improved performance on APR tasks across various bug types and programming language~\cite{rapgen, finetune, finetune2, security, security2}. LLMs are initially pre-trained on extensive unlabeled datasets, and fine-tuning involves further training on a smaller, task-specific dataset. This process intuitively allows LLMs to adjust their weights, enabling them to perform effectively on specific downstream tasks (e.g. code generation, defect detection, and automated program repair).
In the APR community, Jiang et al.\cite{finetune} fine-tune four LLMs on the ARP dataset, showing a 31\% to 1267\% performance improvement over vanilla LLMs.
Similarly, Huang et al.\cite{finetune2} fine-tune five models with more comprehensive strategies, achieving results that surpass state-of-the-art APR tools.

However, their studies demonstrate a limitation that fine-tuning LLMs on limited training instances leads to suboptimal performance. Besides, prior works indicate that the effectiveness of fine-tuning is greatly influenced by the amount of downstream data available~\cite{gu-etal-2022-ppt, finetune, ptr, prompttuning, no, zhang2022differentiable}. In practice, data scarcity is a common issue in APR~\cite{datascarvity, datascarvity1}; the limited amount of training data, compared to the scale of model parameters, may prevent the model from acquiring sufficient knowledge of code syntax and semantics for APR tasks, limiting the performance of fine-tuned LLMs. 


To address the aforementioned limitation of fine-tuning, we adapt prompt tuning for enhanced APR.  
Prompt tuning~\cite{ptr,prompttuning, li-liang-2021-prefix, hard3, liu-etal-2022-ptuning} is recently proposed and initially applied in the field of natural language process (NLP). It trains a pre-trained model using a small amount of task-specific data along with a set of prompts. These prompts provide task-specific knowledge, which is crucial in data scarcity scenarios, to guide the model's adaption to downstream tasks~\cite{shen2021partial,hard3, li-liang-2021-prefix, no, assess}.
Prompt tuning has been proven to achieve superior performance in NLP tasks (e.g. text classification~\cite{ptr}) and some code intelligence tasks~\cite{no, assess, prompt_in_code} (e.g. defect detection, code summarization, and code translation), even in data scarcity scenarios. However, its effectiveness in APR has not been extensively explored. 

In our work, we adapt prompt tuning for enhanced APR and conduct a comprehensive study to evaluate its effectiveness in data scarcity scenarios, using three LLMs of different sizes and six diverse datasets across four programming languages.
In addition to adapting basic prompt tuning for APR, 
our approach also leverages domain knowledge related to bugs during the prompt tuning process.

Prior works have demonstrated the effectiveness of incorporating external knowledge in the prompt tuning process for various NLP tasks, such as relation extraction~\cite{knowprompt} and text classification~\cite{knowledgeable}.
For code intelligence tasks, leveraging domain knowledge has proven to be highly effective across pre-training~\cite{guo-etal-2022-unixcoder}, prompt engineering~\cite{rapgen, typefix} and fine-tuning~\cite{tfix} paradigms. 
However, there is limited research exploring and evaluating the effectiveness of incorporating domain knowledge into the prompt-tuning process specifically for APR tasks.

Our approach addresses this gap by integrating six distinct types of bug- and code-related domain knowledge into specially designed prompt templates during the prompt-tuning process. We select prompt tokens to concatenate specific domain knowledge with the buggy program, forming a comprehensive prompt for the model.
This method, which we refer to as \emph{knowledge prompt tuning}, adapts prompt tuning to optimize both the model and the prompt tokens, allowing the model to learn and apply domain knowledge effectively for APR tasks. 


\begin{figure}
  \centering 
  \includegraphics[width=0.5\textwidth]{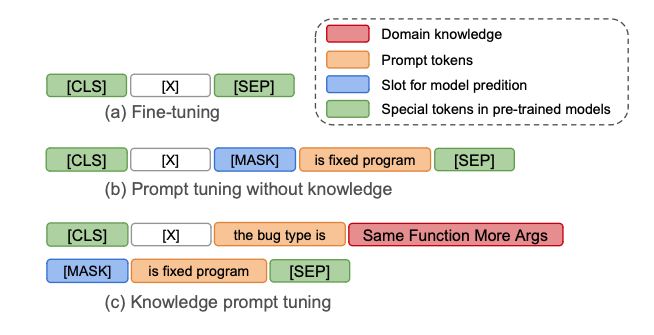}
  \vspace{-25pt}
  \caption{The examples of inputs to CodeT5+ in the paradigm of fine-tuning and prompt tuning. $[X]$ is the slot for buggy code. The orange rectangles represent the prompt tokens, which can be either fixed natural language tokens or learnable soft tokens during prompt tuning.}
  \label{fig1} 
  \vspace{-15pt}
\end{figure}

Fig.~\ref{fig1} illustrates sample inputs to the CodeT5+ model in the paradigm of fine-tuning and prompt tuning with or without domain knowledge.
In this figure, $[X]$ denotes a placeholder for buggy code; a concrete example of buggy code $[X]$ is displayed in Fig.~\ref{fig3}.
Fig.~\ref{fig1}(a) describes how the model integrates a buggy code snippet with special tokens as input to generate fix patches in fine-tuning.
In contrast, Fig.~\ref{fig1}(b) demonstrates that in the process of prompt tuning without domain knowledge, we extend the buggy code snippets with additional prompt tokens, shown in orange rectangles, as input to guide the model toward generating a complete fix patch in the masked token slot.
As a different way, knowledge prompt tuning embeds further domain knowledge that may be specific to the buggy code into the prompts, which are shown in red rectangles in Fig.~\ref{fig1}(c), combining them with the buggy code and additional prompt tokens.



Our main contributions can be summarized as follows:

\begin{enumerate}[(1),nosep,leftmargin=1.5em]
\item To the best of our knowledge, our work is the first to adapt prompt tuning to the APR task, utilize domain knowledge in prompt tuning, and extensively evaluate its effectiveness in APR against fine-tuning methods in data scarcity scenarios.


\item We investigate the influence of different prompts, domain knowledge, and sizes of training datasets on the performance of prompt tuning. 

\item Based on our findings, we explore the implications of utilizing domain knowledge and various prompts in prompt tuning, and we propose future directions for research in this area.

\end{enumerate}

%% file: background.tex
\subsection{Fine-tuning}
Fine-tuning~\cite{fin1, fin2, fin3, finetune, finetune2} is a machine learning technique used to adapt a pre-trained model to perform a specific task or set of tasks. In fine-tuning, a model that has been initially trained on a large and diverse dataset, often referred to as a "pre-trained" model, is further trained on a smaller, task-specific dataset. The goal is to modify the pre-trained model's parameters to make it proficient in solving a particular problem without completely overwriting the valuable knowledge it gained during its initial pre-training.

Fine-tuning typically involves adjusting the model's weights and parameters through additional training iterations using the task-specific dataset. More specifically, suppose a dataset contains samples denoted as $X$ and corresponding labels denoted as $Y$, the fine-tuning process is to maximize the likelihood of predicting $Y$ given $X$, $Pr_{\theta}(Y|X)$.

\subsection{Prompt tuning}
Compared to fine-tuning, prompt tuning aims to provide more guidance to the model by incorporating additional tokens, which are denoted as prompt, as input to the model, and tune the prompts with the model on a small set of data. In detail, suppose we have a buggy source code snippet as a series of tokens $X$ and the corresponding corrected program as a series of tokens $Y$. In prompt tuning, besides the input tokens $X$, we also add additional information for the model, which is a series of prompt tokens, denoted as $P$. The goal is to maximize the likelihood of the correct $Y$, $Pr_{\theta}(Y|X; P)$ by optimizing the values of both $\theta$ and $P$. 

\begin{figure}[t]
  \centering 
  \includegraphics[width=0.85\columnwidth]{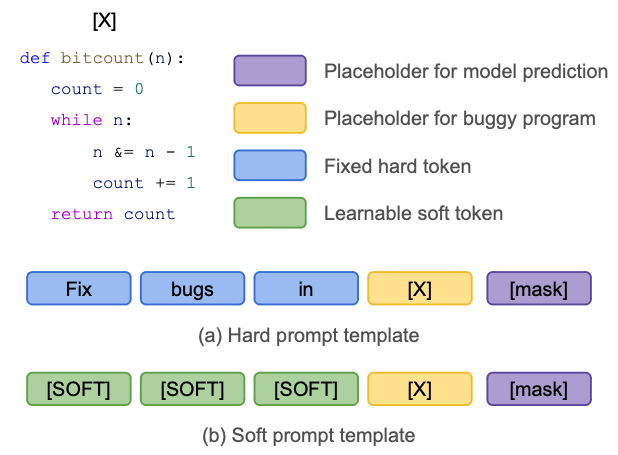} 
  \vspace{-10pt}  
  \caption{ Illustration on hard prompt and soft prompt, where [X] and [mask] indicate the input slot and output slot respectively.} 
  \label{fig3} 
  \vspace{-13pt}
\end{figure}

In this work, we adapt prompt tuning for automated program repair.
We provide an example in Fig.~\ref{fig3} to illustrate the inputs used in the tuning of a code model (e.g., CodeT5+ \cite{wang-etal-2023-codet5} in our experiments).
The buggy source code (represented as [X] in the figure) is part of the input prompts fed to the model, which would auto-regressively generate the output as fix patches (represented as [mask] in the figure).
Prompt tuning creates a template to wrap the buggy source code with additional information as input to the model. This template could consist of natural language tokens that remain unchanged during prompt tuning
(e.g., ``Fix'', ``bugs'', ``in'', in Fig.~\ref{fig3}(a)), 
or it could include soft tokens that are optimized during the tuning process (e.g. the three ``[SOFT]'' in Fig.~\ref{fig3}(b)).
In our experiments, we feed such prompts along with the actual buggy source code as input to the model and aim for the model to predict the output in the ``[mask]'' position, which ideally should be the complete fix patch for the buggy code. 

Depending on the tokens used in the prompts, researchers have categorized prompt tuning into two types: hard prompts (a.k.a.\ discrete prompts) and soft prompts (a.k.a.\ continuous prompts),
as illustrated in Fig~\ref{fig3}(a) and (b).

\subsubsection{Hard Prompt}
The hard prompt~\cite{gu-etal-2022-ppt, ptr, hard3} is a method to rewrite the model's input by introducing a series of natural language tokens as prompts. Each token added to the model's input is human-understandable~\cite{gu-etal-2022-ppt}.

For example, when adapting prompt tuning to APR, we may design the prompt template that can be expressed as follows:
\begin{equation}
{\displaystyle f(X,mask) = \text{ Fix the bug in } [X] [mask]}
\end{equation}
where $[X]$ is a placeholder for the input slot for the buggy source program and $[mask]$ is a placeholder for the output slot for the generated fix patch. The tuning objective of the model is to predict the repaired program at the output slot based on the input buggy program. 


\subsubsection{Soft Prompt}
\label{sec:sp} 
To alleviate the inflexibility of hard prompts, researchers propose an alternative prompt tuning method called soft prompt~\cite{ptr,li-liang-2021-prefix, soft3}. In contrast to a hard prompt, the tokens within the soft prompt template are not static discrete words from natural language. Instead, these tokens are continuous vectors that can be learned during the tuning phase. The soft tokens are typically vectors and not human-interpretable. 

Soft prompts can be obtained by simply replacing a hard prompt token with a soft token, denoted as [SOFT]. The embedding of the soft token can be learned during tuning. An example of a Soft Prompt is:
\begin{equation}
{\displaystyle    f(X,mask) = [X] \text{ [SOFT] [SOFT] [SOFT] } [mask]}
\end{equation}


%% file: ourapp.tex
To adapt basic and knowledge prompt tuning in APR, we design prompt templates that include hard and soft variants for basic and knowledge prompts.
(cf.\ Section \ref{sec:background}). 
 
\vspace{-3pt}
\subsection{Basic Prompt Tuning}
\label{sec:bp}
We refer to prompts that do not contain any additional code- or bug-related domain knowledge beyond the buggy programs and masked targets as \emph{basic prompts (BP)}, and the prompt tuning process using these prompts as \emph{basic prompt tuning}. 

As shown in Table~\ref{tab:1}, we design seven distinct prompt templates for basic prompts, extending previous work to enable a comprehensive exploration of how different prompt templates influence the prompt-tuning process. These templates are labeled from BP1 to BP7.
[X] and [mask] serve as placeholders for buggy programs and model predictions, respectively. These designs are based on variations in the positions of prompt tokens, 
[X], and [mask], ensuring that the prompt templates are semantically coherent and meaningful.

For each prompt template, we design both hard and soft variants to further investigate how different prompt token types influence the prompt tuning process. 
\emph{Hard basic prompts (HBP)} are constructed by concatenating [X] and [mask] with various fixed, human-readable tokens, forming structured and interpretable prompts. 
By replacing the hard tokens in the HBP with soft tokens, we derive \emph{soft basic prompts (SBP)}. Based on how the soft tokens are initialized, we classify SBP into two versions: 
\emph{SBP$_{\text{initialized}}$}, where the soft tokens are initialized by the corresponding hard tokens in HBP (e.g. for SBP1$_{\text{initialized}}$, three [SOFT] tokens are initialized by ``is'' ``fixed'' ``program'') and \emph{SBP$_{\text{random}}$}, where the soft tokens are initialized randomly.

\begin{table}
\footnotesize
  \centering
  \caption{Templates of hard and soft basic prompts}
  \label{tab:1}
  \setlength{\arrayrulewidth}{0.1pt}
  \renewcommand{\arraystretch}{1.3}
  \begin{tabular}{m{0.1cm}m{4cm}|m{3.4cm}}
    \toprule 
    & \textbf{Basic Hard prompt} &  \textbf{Basic Soft prompt}  \\
    \midrule 
     1 & \text{[X] [mask] is fixed program} & \text{[X] [mask] [SOFT] * 3}\\
    \midrule 
     2 &\text{[X] fixed program is [mask]}  & \text{[X] [SOFT] * 3 [mask]} \\
    \midrule 
     3 &\text{Fix bug in [X] [mask]}  & \text{[SOFT] * 3 [X] [mask]} \\
    \midrule 
   4 &\text{Fix [X] fixed program is [mask]}  & \text{[SOFT] [X] [SOFT] * 3 [mask]} \\
    \midrule 
    5 &\text{Fix [X] [mask] is fixed program}  & \text{[SOFT] [X] [mask] [SOFT] * 3 } \\
    \midrule 
    6 &\text{[X]} is buggy program \text{[mask]} is fixed program & \text{[X] [SOFT] * 3 } \text{[mask] [SOFT] * 3 }\\
    \midrule 
    7 & Fix \text{[X]} is buggy program \text{[mask]} is fixed program  & \text{[SOFT] [X] [SOFT] * 3} \text{[mask] [SOFT] * 3 } \\
    \bottomrule
  \end{tabular}
\setlength{\arrayrulewidth}{0.4pt}
\vspace{-15pt}
\renewcommand{\arraystretch}{1}
\end{table}


\vspace{-3pt}
\subsection{Knowledge Prompt Tuning}
\label{sec:kp}
We define \emph{knowledge prompts (KP)} as prompts that incorporate additional code- or bug-related knowledge specific to each buggy program, beyond the buggy code itself and the generic hard or soft tokens used in BPs. 
Unlike BP, which includes only generic hard or soft tokens apart from [X] and [mask], KP leverages detailed domain knowledge, such as error messages, abstract syntax tree (AST) of buggy nodes, or specific bug types, to provide a richer context. 
We extract six kinds of domain knowledge from metadata provided in our evaluation dataset, as summarized in Table~\ref{tab:knowledge}.

In Table~\ref{tab:def}, we present a few examples of our designed KP templates for each type of domain knowledge and model.
These prompt templates are designed to incorporate various types of domain knowledge, with the aim of making the entire prompt more readable and coherent by adding some generic tokens. 
For example, to incorporate repair actions as domain knowledge, into KPs, we add generic tokens ``by'' ``taking'' ``repair'' ``action'' to help the model better understand the context. 

Moreover, if a dataset provides more than one type of domain knowledge, we design additional KP templates by incorporating multiple domain knowledge types.
The generic prompt tokens (i.e., tokens other than [X], domain knowledge, [mask]) in each KP template may be either hard or soft tokens, resulting in what are respectively called soft KPs and hard KPs. 

\noindent
\textbf{Model Usage: Generative vs.\ Infilling}:
Considering the differences between generative and infilling models during the inference phase, we create distinct KP templates for each model type, as listed in Table~\ref{tab:def}. 
For infilling models, we position [mask] within the middle of the template, rather than at the end, to encourage the model to generate fixed patches by utilizing both preceding and following context, whereas for generative models, [mask] is placed at the end of the template since the generative models predict the next token based on the previous tokens only.

\begin{table}
  \footnotesize
  \centering
  \caption{Domain knowledge incorporated in knowledge prompts for each dataset}
  \label{tab:knowledge}
  \setlength{\arrayrulewidth}{0.1pt}
  \renewcommand{\arraystretch}{1}
  \begin{tabular}{l p{5cm}} 
    \toprule 
     \textbf{Dataset} &  \textbf{Domain knowledge}  \\
    \midrule 
     Defects4j~\cite{defects4j} & Repair action, Repair pattern\\
     ManySStuBs4J-SStuBs~\cite{manysstubs} & Bug type, AST of buggy nodes \\
     TFix~\cite{tfix}& Bug type, Error message \\
     xCodeEval~\cite{xcodeeval} & Error message, Tags of potential algorithmic techniques in buggy code\\
    \bottomrule
  \end{tabular}
  \vspace{-15pt}
\end{table}

\begin{table*}
  \centering
  \scriptsize
  \caption{Examples of knowledge prompt templates with various domain knowledge}
  \label{tab:def}
  \renewcommand{\arraystretch}{1}
  \begin{tabular}{c p{16cm}} 
    \toprule 
    & \multicolumn{1}{c}{\textbf{Knowledge prompt templates for infilling models}} \\
        \midrule 
1  & Please fix a buggy program \text{[X]} the bug type is \text{[bugType]} \text{[mask]} is the fixed version  \\
2  & Please fix a buggy program \text{[X]} by taking repair actions \text{[repairAction]} \text{[mask]} is the fixed version  \\
3  & Please fix a buggy program \text{[X]} by following repair patterns \text{[repairPattern]} \text{[mask]} is the fixed version \\
    \midrule 
& \multicolumn{1}{c}{\textbf{Knowledge prompt templates for generative models}} \\
    \midrule 
1  & Please fix a buggy program \text{[X]} the bug type is \text{[bugType]} the fixed version is [mask] \\
2  & Please fix a buggy program \text{[X]} by taking repair actions \text{[repairAction]} the fixed version is \text{[mask]} \\
3  & Please fix a buggy program \text{[X]} by following repair patterns \text{[repairPattern]} the fixed version is \text{[mask]} \\

\bottomrule
  \end{tabular}
  \vspace{-15pt}
\end{table*}

\vspace{-5pt}
\subsection{Prompt Tuning and Fine-tuning Implementations}
\label{sec:openprompt}
We use OpenPrompt~\cite{openprompt} to construct prompts for prompt-tuning and PyTorch as the framework to generate experimental results for both prompt-tuning and fine-tuning. OpenPrompt is a flexible toolkit for prompt-based learning, enabling the design and application of custom prompt templates with both fixed hard tokens and learnable soft tokens. Its versatility supports our knowledge prompt tuning strategy. However, OpenPrompt has not been actively maintained for over a year, limiting access to the latest models for prompt tuning. For fair comparison, we use fine-tuning, which updates all model parameters, as our baseline, aligning with the full parameter updates required by OpenPrompt in prompt tuning.

%% file: experimental.tex
\begin{figure}
  \centering 
  \includegraphics[width=0.5\textwidth]{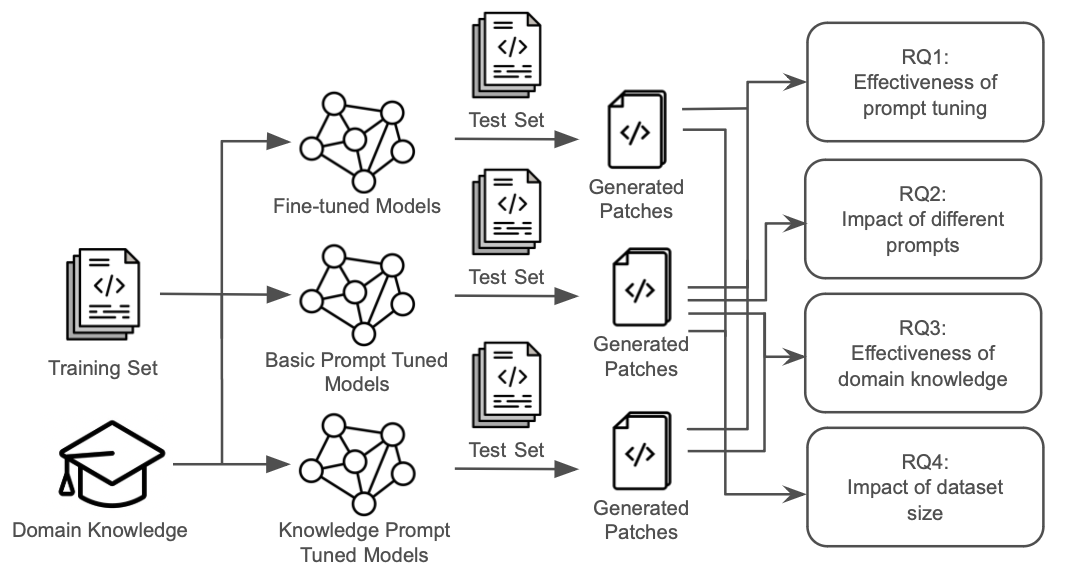} 
    \vspace{-15pt}
  \caption{Overview of experimental design} 
  \label{fig_overview} 
  \vspace{-15pt}
\end{figure}

Fig.~\ref{fig_overview} shows the overview of our experimental design.
We ask four research questions (RQs) in relation to the evaluation of the effectiveness of prompt tuning across various scenarios.

We select three state-of-the-art models as our base models to conduct our experiments and evaluate prompt tuning on six APR benchmarks across four programming languages. To ensure the experiment results are as fair as possible, we utilize three popular evaluation metrics in APR.


\subsection{Research Questions}
In this paper, we aim to find out the answers to the following research questions through comprehensive experiments:

\textbf{RQ1}: How effective is prompt tuning in completing APR tasks in data scarcity scenarios?

\textbf{RQ2}: How do different types of prompts impact the performance of prompt tuning?

\textbf{RQ3}: To what extent does knowledge prompt benefit from domain knowledge integration?

\textbf{RQ4}: What is the impact of dataset size on prompt tuning performance?


\vspace{-2pt}

\subsection{Studied models}
To conduct an in-depth evaluation of prompt tuning, we select pre-trained models based on several criteria: (1) the models must be pre-trained on a code corpus and open-source for accessibility; (2) models should have fewer than 1.5 billion parameters due to computational constraints, excluding larger models like GPT-J~\cite{gpt-j} and CodeLLaMa~\cite{codellama}; (3) the models should directly apply to APR without requiring architectural modifications or speciall format of input, leading to the exclusion of encoder-only models like CodeBERT. We selected three state-of-the-art models for generation tasks.

\textbf{CodeT5+} is an advanced version of CodeT5, excelling in a wide range of code intelligence tasks. It is pre-trained on CodeSearchNet and additional datasets from GitHub with the objectives of span denoising and causal language modeling, enabling strong performance in both mask prediction and Seq2Seq generation tasks. In our experiments, we use two versions: \textbf{CodeT5+ 220M} (220 million parameters) and \textbf{CodeT5+ 770M} (770 million parameters).

\textbf{GPT-Neo}~\cite{gpt-neo} is an open-source auto-regressive model based on the Transformer architecture. Pre-trained on the Pile dataset, which includes diverse texts and code, it is highly effective for generation tasks. We use \textbf{GPT-Neo 1.3B} (1.3 billion parameters) in our experiments.

\vspace{-2pt}
\subsection{Evaluation Datasets}
\vspace{-2pt}
\label{sec:datasets}
 





We choose six commonly used APR benchmarks from the literature across four programming languages.
Due to the model's maximum input length restriction, we treat each single-hunk (continuous lines) fix as a separate instance for each dataset extracted from real-world projects, then we split the data into training, validation, and test sets in an 8:1:1 ratio for each dataset for both fine-tuning and prompt tuning experiments. 
However, the six datasets vary greatly in size, with some containing tens of thousands of instances and others having fewer than a thousand. To focus on data scarcity scenarios, we scale all datasets to a similar size by sampling 1\% from the larger ones, while keeping the smaller datasets unchanged.
To reduce the effect of randomness of sampling on the results, we generate three subsets for each of the sampled datasets using three distinct seeds and the final experimental results present in this work are the average of the results from these three subsets. The statistics of our evaluation datasets after sampling are listed in Table~\ref{tab:dataset}. 

\begin{table}
  \centering
  \caption{Statistic of selected evaluation datasets}
  \scriptsize
  \label{tab:dataset}
  \begin{tabular}{l|c|ccc}
    \toprule 
     \multirow{2}{*}{\textbf{Dataset}} & \multirow{2}{*}{\textbf{\textbf{Language}}} & \textbf{Training} & \textbf{Val.} & \textbf{Test} \\
     && \textbf{Set} & \textbf{Set} & \textbf{Set} \\
    \midrule 
    BugsInPy~\cite{bugsinpy} & Python & 932 & 112 & 122 \\
    
    Code Refinement~\cite{coderefine} & Java & 523 & 65 & 65 \\

    Defects4J & Java & 502 & 63 & 64 \\

    ManySStuBs4J-SStuBs & Java & 260 & 32 & 32 \\

    TFix & JavaScript & 816 & 102 & 102 \\
    
    xCodeEval-APR-C & C & 1,082 & 135 & 135 \\

    \bottomrule
  \end{tabular}
  \vspace{-15pt}
\end{table}


\subsection{Evaluation Metrics}
\label{sec:metrics}
To compare the repair performance of various models, we choose three evaluation metrics commonly used in the literature~\cite{tfix, no}:
\footnote{Since most of the datasets do not have test cases, we do not use test cases as an evaluation metric although they can be effective in checking the functional correctness of repairs.}

\textit{A) Exact Match (EM):}
\textit{EM} refers to the generated patches that exactly match the fix reference. In this work, our results are present in EM rate, the percentage of EM instances across the test set, except for the results of RQ4. 

\textit{B) Syntactically Correct Patch (SC):} Since some buggy programs may have more than one correct fix, in addition to EM, we define the generated patches that are syntactically equivalent to the fixed reference as syntactically correct patches by comparing their syntax tree. 

\textit{C) CodeBLEU ~\cite{ren2020codebleu}: } 
In addition to these two metrics, EM and SC, a more relaxed metric, \textit{CodeBLEU}, is included to measure the extent to which a program is repaired. Unlike the traditional BLEU score,
CodeBLEU more accurately measures the similarity between generated patches and reference fixes by taking both syntax and semantics of programs into account. 
%


\subsection{Implementation Details}
\vspace{-1pt}
\label{sec:implementation}
In our experiments, all pre-trained LLMs are loaded from the official versions available on Hugging Face. Details regarding the hyperparameter setting during training and generation processes are provided in Table~\ref{tab:hyperparameter}. The experiments on CodeT5+ models and GPT-Neo models were conducted on a server equipped with an NVIDIA A40 48GB GPU and H100 80GB GPU, respectively.
\begin{table}[h]
\vspace{-10pt}
  \scriptsize
  \centering
  \caption{Hyperparameter Settings}
  \label{tab:hyperparameter}
  \setlength{\arrayrulewidth}{0.1pt}
  \renewcommand{\arraystretch}{1}
  \begin{tabular}{c|c||c|c} 
    \toprule 
     Training Hyp. & Value & Generation Hyp. & Value
      \\
    \midrule 
Optimizer & AdamW & Temperature & 1.0\\
Adam Epsilon   & 1e-8 &  Sample & False\\
Initial Learning Rate & 5e-5 & Repetition Penalty & 1.0 \\
LR scheduler  & Linear & Top p & 0.9 \\
Training epochs  & 10 &  Bean Number & 5\\
    \bottomrule
  \end{tabular}

  \vspace{-13pt}
\end{table}

%% file: result.tex
\subsection{RQ1: Effectiveness of Prompt Tuning}

\begin{table*}
\scriptsize
  \centering
  \caption{Results of prompt tuning and fine-tuning on six datasets}
  \label{tab:6}
  \renewcommand{\arraystretch}{1}
  \begin{tabular}{c c|c c c|c c c|c c c} 
    \toprule 
     \multirow{2}{*}{\textbf{Model}} & \multirow{2}{*}{\textbf{\textbf{Tuning Methods}}} & \multicolumn{3}{c|}{\textbf{BugsInPy}}  & \multicolumn{3}{c|}{\textbf{Code Refinement}}  & \multicolumn{3}{c}{\textbf{Defects4J}}    \\
     && EM & SC & CodeBLEU & EM & SC & CodeBLEU& EM & SC & CodeBLEU  \\
    \midrule 

    Naive Copy & -- & 0 & 0 & 76.94 & 0 & 0 & 75.82 & 0 & 0 & 75.94 \\\hline
    \multirow{2}{*}{CodeT5+ 220M} & Fine-tune  & 13.93 & 14.75 & 77.66 & 0.51 & 0.51 & 75.45 & 5.21 & 5.21 & 76.22 \\
    & prompt tuning  & 15.57 & 16.39 & 76.65 & 1.54 & 1.54 & 86.42 & \textbf{14.06} & 14.06 & 75.41 \\\midrule 
    \multirow{2}{*}{CodeT5+ 770M} & Fine-tune  & 13.52 & 13.93 &76.95& 0 & 0 & 75.41 & 4.17 & 4.17 & 75.98 \\
    & prompt tuning  & \textbf{18.03} & 18.85 & 77.17 & \textbf{2.56} & 2.56 & 86.27 & 12.50 & 12.50 & 75.45 \\\midrule 
    \multirow{2}{*}{GPT-Neo 1.3B} & Fine-tune  & 0 & 0 &75.79 & 0 & 0 & 75.82 & 0 & 0 & 73.7\\
    & prompt tuning  & 13.11 & 14.75 & 74.38 & 1.03 & 1.54 & 84.16 & 12.50 & 12.50 & 73.87\\
    \midrule 
    \midrule
    
     \multirow{2}{*}{\textbf{Model}} & \multirow{2}{*}{\textbf{\textbf{Tuning Methods}}} &  \multicolumn{3}{c|}{\textbf{ManySStuBs4J}}  & \multicolumn{3}{c|}{\textbf{TFix}}  & \multicolumn{3}{c}{\textbf{xCodeEval}}   \\
     && EM & SC & CodeBLEU &EM & SC & CodeBLEU & EM & SC & CodeBLEU   \\
    \midrule 
    
    Naive Copy & -- & 0 & 0 & 91.10 & 0 & 0 & 57.19 & 0 & 0 & 75.45\\\hline
    \multirow{2}{*}{CodeT5+ 220M} & Fine-tune & 49.31 & 49.31 & 94.92 & 20.92 & 21.90 & 63.63 & 1.60 & 1.73 & 73.53 \\
    & prompt tuning  & \textbf{57.29} & 57.29 & 94.13 & 23.53 & 24.18 & 64.53 & 3.46 & 3.46 & 75.21 \\\midrule 
    \multirow{2}{*}{CodeT5+ 770M} & Fine-tune  & 48.61 & 48.61 & 94.17 & 21.13 & 22.22 & 64.16 & 2.35 & 2.35 & 73.56 \\
    & prompt tuning & 57.29 & 57.29 & 93.69 & \textbf{25.82} & 26.14 & 66.08 & \textbf{4.44} &4.69 & 74.92\\\midrule 
    \multirow{2}{*}{GPT-Neo 1.3B} & Fine-tune  & 0 & 0 & 90.36 & 0 & 0 & 56.40 & 0 & 0 & 74.89 \\
    & prompt tuning  & 54.17 & 54.17 & 93.53 & 17.97 & 17.97 & 62.47 & 2.22 & 2.22 & 74.98 \\
    \bottomrule
  \end{tabular}
\vspace{-15pt}
\end{table*}

In this section, we explore the effectiveness of prompt tuning by comparing the performance of prompt-tuned models with the performance of fine-tuned baselines on six selected APR datasets with limited training instances. Our experimental setup is in scenarios of data scarcity, as the sizes of our training sets range from 260 to 1,082 across different datasets, which are relatively small compared to other works in fine-tuning LLMs~\cite{finetune, finetune2}. 
The results we present in Table~\ref{tab:6} are the best performance we achieved across various basic and knowledge prompts we introduced in Table~\ref{sec:ourapp}.

First of all, we notice that ``Naive Copy'', which simply copies buggy code as fix patches, yields high CodeBLEU scores but zero EM or SC, indicating significant overlap between buggy code and its fix. Therefore, EM is prioritized as the primary metric. 
Overall, for two CodeT5+ models, prompt tuning achieves performance improvements ranging from 11.63\% to 201.96\% and 87.33\% on average across various datasets and base models, while  GPT-Neo 1.3B demonstrates significant improvement with prompt tuning compared to fine-tuning, the EM rate increases from 0\% to values ranging between 1.03\% and 54.17\% across various datasets. The results highlight the superior effectiveness of prompt tuning over fine-tuning.

\subsubsection{Performance comparison across datasets}
Looking into individual datasets reveals that prompt tuning consistently outperforms fine-tuning.

With ManySStuBs4J, our models achieved the highest EM rate in both prompt tuning and fine-tuning methods across all six datasets and three models. With ManySStuBs4J, fine-tuned models achieve EM rates of 49.31\% and 48.61\% for CodeT5+ 220M and CodeT5+ 770M, respectively. This notable performance is attributed to the fact that the bugs in ManySStuBs4J are typically simple single-statement bugs, which also contributes to the high CodeBLEU scores. Nevertheless, prompt tuning manages to further enhance the models' performance, achieving EM rates of 57.29\% for both models. For GPT-Neo, prompt tuning achieves a 54.17\% EM rate, compared to 0\% with fine-tuning.

In contrast, with Code Refinement, although prompt tuning further boosts the models' performance compared to fine-tuning, the results of this dataset are less remarkable than those of others. We suspect this may be caused by the variable anonymization in the source code of Code Refinement, which limits the model's ability to learn semantic information from variable and function names. In Code Refinement, variable and method names are abstracted into generic identifiers (e.g., VAR\_1, METHOD\_1), making it harder to understand the context. Without explicit variable names, the model struggles to infer the code's function through semantics. In this case, generating fixed patches requires the model to learn more about the syntax knowledge of the given program, which becomes challenging when data is scarce.

\vspace{2pt}
\noindent\colorbox{gray!30}{%
\parbox{\dimexpr\linewidth-2\fboxsep\relax}{%
\textbf{Finding 1}: In data scarcity scenarios, prompt tuning consistently outperforms fine-tuning in APR tasks, with respect to four selected LLMs and six selected datasets across four programming languages. 
}%
}
\vspace{-3pt}


\subsubsection{Performance comparison across models}
Across the three selected models, we find that for each model, prompt-tuned models consistently outperform fine-tuned models. 

The improvement of prompt tuning over fine-tuning is most pronounced for GPT-Neo, compared to CodeT5+ 220M and 770M. For each dataset, fine-tuned GPT-Neo 1.3B underperforms the fine-tuned CodeT5 models, with its EM rates remaining at 0. This is largely due to GPT-Neo’s broader and less task-specific pre-training process, which results in lower performance during fine-tuning without explicit instruction, particularly in data scarcity scenarios.
On the other hand, CodeT5+ is pre-trained to handle various code generation tasks, making it adapt better to APR data during fine-tuning, even with limited training data.
Nevertheless, through prompt tuning, GPT-Neo 1.3B can achieve performance comparable to CodeT5+, as the prompt tokens in prompt tuning provide task-specific guidance that instructs GPT-Neo 1.3B to perform APR tasks, in contrast to the unguided fine-tuning process. This demonstrates that in limited data situations, prompt tuning leads to a greater improvement for GPT-Neo compared to fine-tuning, than it does for CodeT5+ models.

\vspace{1pt}
\noindent\colorbox{gray!30}{%
\parbox{\dimexpr\linewidth-2\fboxsep\relax}{%
\textbf{Finding 2}: In data scarcity scenarios, prompt tuning boosts performance across all three models compared to fine-tuning, with GPT-Neo showing greater improvement than CodeT5+ models.
}%
}
\vspace{-8pt}

\begin{table}
\scriptsize
  \centering
  \caption{Comparison of Prompt Tuning Results Using Hard and Soft Basic Prompts}
  \label{tab:different_types}
  \renewcommand{\arraystretch}{1}
  \begin{tabular}{l|c|ccc} 
    \toprule 
    \textbf{Dataset} & \textbf{Model} & HBP & SBP$_{initialized}$ & SBP$_{random}$ \\
    \midrule 
    \multirow{3}{*}{BugsInPy} & CodeT5+ 220M & 14.75 & \textbf{15.57} & 13.93 \\
                                       & CodeT5+ 770M & 15.57 & \textbf{18.03} & 15.57 \\
                                       & GPT-Neo 1.3B  & \textbf{13.11} & \textbf{13.11} & -- \\
    \midrule 
    \multirow{3}{*}{Defects4J} & CodeT5+ 220M & 9.38 & 10.94 & \textbf{14.06} \\
                                       & CodeT5+ 770M & 10.94 & \textbf{12.50} & \textbf{12.50} \\
                                       & GPT-Neo 1.3B  & 10.94 & \textbf{12.50} & -- \\
    \midrule 
    \multirow{3}{*}{ManySStuBs4J} & CodeT5+ 220M & 56.25 & \textbf{57.29} & 55.21 \\
                                       & CodeT5+ 770M & 56.25 & \textbf{57.29} & \textbf{57.29} \\
                                       & GPT-Neo 1.3B  & 53.13 & \textbf{54.17} & -- \\
    \midrule 
    \multirow{3}{*}{TFix} & CodeT5+ 220M & 23.20 & \textbf{23.53} & 23.20 \\
                                       & CodeT5+ 770M & 22.55 & 25.49 & \textbf{25.82} \\
                                       & GPT-Neo 1.3B  & 16.34 & \textbf{17.97} & -- \\
    \midrule 
    \multirow{3}{*}{xCodeEval} & CodeT5+ 220M & 2.96 & \textbf{3.46} & \textbf{3.46} \\
                                       & CodeT5+ 770M & 4.20 & 3.95 & \textbf{4.44} \\
                                       & GPT-Neo 1.3B  & 1.98 & \textbf{2.22} & -- \\

    \bottomrule
  \end{tabular}
\vspace{-18pt}
\end{table}

\begin{table*}
\scriptsize
  \centering
  \caption{Comparison of prompt tuning across different basic prompt templates on four datasets}
  \label{tab:different_design}
  \renewcommand{\arraystretch}{1}
  \begin{tabular}{c|ccc|ccc|ccc|ccc} 
    \toprule 
    \textbf{Basic} & \multicolumn{3}{c|}{\textbf{BugsInPy}} &  \multicolumn{3}{c}{\textbf{ManySStuBs4J}}  & \multicolumn{3}{c}{\textbf{TFix}}  &  \multicolumn{3}{c}{\textbf{xCodeEval}}\\
    \textbf{Prompt} & CodeT5+& CodeT5+  & GPT-Neo  & CodeT5+  & CodeT5+  & GPT-Neo  & CodeT5+ & CodeT5+  & GPT-Neo  & CodeT5+ & CodeT5+  & GPT-Neo\\
    \textbf{Template} & 220M & 770M & 1.3B & 220M & 770M & 1.3B & 220M & 770M & 1.3B & 220M & 770M & 1.3B\\
    \midrule 
    1 & 13.93 & 15.57 & 4.10 & \textbf{57.29} & 56.25 & 1.04 & 22.88 & 22.22 & 6.54 & 3.21 & 3.70&  0.74\\
    2 & 13.11 & 12.30 & 12.30 & 55.21 & 54.17 & \textbf{54.17} & \textbf{23.20} & 21.90 & \textbf{17.97}&2.72 & 4.20 & \textbf{1.98} \\
    3 & 13.93 & 13.93 & 13.11 & 56.25 & 56.25 & 51.04 & 21.57 & 22.22 & 16.67& 2.96 &3.95 & 1.98\\
    4 & 13.93 & \textbf{18.03} & \textbf{13.93} & 55.21 & \textbf{57.29} & 52.08 & 22.22 & \textbf{23.86} & 16.99& 2.47 & 3.70& 1.23\\
    5 & 13.93 & 16.39 & 3.28 & 57.29 & 55.21 & 8.33 & 22.22 & 21.90 & 6.54 & 3.46 & 2.96 & 0.74\\
    6 & \textbf{15.57} & 15.57 & 4.10 & 55.21 & 55.21 & 13.54 & 21.90 & 20.92 & 11.76 &3.21 &\textbf{4.44} & 0.74\\
    7 & 13.93 & 14.75 & 4.92 & 55.21 & 56.25 & 22.92 & 21.90 & 21.90 & 11.44 &\textbf{3.46} & 3.46 & 1.48\\
    \bottomrule
  \end{tabular}
  \vspace{-15pt}
\end{table*}
\subsection{RQ2: Impact of Different Basic Prompts}
\label{sec:difprompt}

In this section, we investigate the impact of different basic prompts. First, we compare the performance of hard basic prompts (HBP), soft basic prompts (SBP) with soft tokens initialized by hard tokens in HBP (SBP$_{\text{initialized}}$) and soft basic prompts (SBP) with soft tokens initialized randomly (SBP$_{\text{random}}$), as shown in Table~\ref{tab:different_types}. Moreover, we compare the different basic prompt templates listed in Table~\ref{tab:1}, with the results shown in Table~\ref{tab:different_design}. Due to space limitations, we present the results from four datasets, each representing a distinct programming language, and we only present the EM rate of each BPs in this section. 

\subsubsection{Comparison of different types of prompts}
\label{sec:hardvssoft}
We have introduced hard basic prompts (HBP) which concatinate [X] and [mask] with natural language tokens, and soft basic prompts (SBP) which concatenate [X] and [mask] with learnable [SOFT] tokens. SBP can be classified into two versions based on how soft tokens are initialized: SBP$_{\text{initialized}}$ 
and SBP$_{\text{random}}$.  

In Table~\ref{tab:different_types}, we compare the performance of HBP, SBP$_{\text{initialized}}$ and  SBP$_{\text{random}}$ across various base models and datasets. GPT-Neo 1.3B fails to converge within the specified number of epochs with SBP$_{\text{random}}$ in our experiment setups. We suspect this is due to its significantly larger number of parameters, which may require more time and data for convergence.
In general, we observe that SBPs are more effective than HBPs in prompt tuning. This conclusion is consistent with findings mentioned in previous works~\cite{no, assess} that hard prompts perform better than soft prompts in classification tasks, but this advantage tends to be weakened in generation tasks. 

By comparing SBP$_{\text{initialized}}$ with SBP$_{\text{random}}$, we observe that in most cases, their performance is quite close to each other, with a difference of less than 1\%. However, there are instances where their performance diverges. For example, in the BugsInPy dataset, CodeT5+ 220M and 770M perform significantly better with SBP$_{\text{initialized}}$ (15.57\% and 18.03\%, respectively) compared to SBP$_{\text{random}}$ (13.93\% and 15.57\%). In such cases, vocabulary initialization tends to yield better results because it offers a more stable and efficient starting point for learning. However, in the Defects4J dataset, CodeT5+ 220M achieves a 3.12\% higher EM rate with SBP$_{\text{random}}$ compared to SBP$_{\text{initialized}}$. This exception could be attributed to random initialization providing a more diverse starting point, which enables broader exploration and may lead the model to discover alternative local optima. The choice between SBP$_{\text{initialized}}$ and SBP$_{\text{random}}$ for prompt tuning should based on the specific task, model, and dataset.

\vspace{2pt}
\noindent\colorbox{gray!30}{%
\parbox{\dimexpr\linewidth-2\fboxsep\relax}{%
\textbf{Finding 3}: Soft basic prompts (SBP) generally outperform hard basic prompts (HBP) in prompt tuning. While the performance of SBP$_{\text{initialized}}$ and SBP$_{\text{random}}$ is typically similar, one can exhibit an advantage over the other depending on the different model and dataset.
}%
}
\vspace{-2pt}

\subsubsection{Comparison of different basic prompt templates}
As mentioned in Table~\ref{tab:1}, 
we create seven distinct basic prompt templates by altering the sequence of [X], [mask], and the prompt tokens. These templates are labeled from BP1 to BP7. The results of seven basic prompt templates are listed in Table~\ref{tab:different_design}. 

The performance of each prompt template in prompt tuning varies across datasets.
Overall, BP4 tends to perform well for CodeT5+ 770M, while BP2 delivers the best results for GPT-Neo 1.3B across most datasets.
For CodeT5+ 220M, the performance gap between the best and worst BP templates ranges from 0.99\% to 2.46\%, while for CodeT5+ 770M, this range is from 1.24\% to 5.73\%. However, when using GPT-Neo 1.3B on the ManySStuBs4J dataset, there is a significant difference in performance across the templates. For example, BP2 achieves the highest EM rate at 54.17\%, while BP1 yields a considerably lower EM rate of 1.04\%. This significant gap is also observed in the results of other datasets. After analyzing the performance of the seven distinct templates, we find that when GPT-Neo is used as the base model, BP1, BP5, BP6, and BP7 consistently underperform, whereas BP2, BP3, and BP4 demonstrate 
better
results, with a pronounced distinction in performance between these two groups.

The performance gap arises because, in BP1, BP5, BP6, and BP7, the [mask] token is placed in the middle of the prompt sequence, with additional prompt tokens following it. Since GPT-Neo 1.3B, as a generative model, determines the next token based on the preceding tokens only, this arrangement results in the prompt tokens after the [mask] token not being effectively utilized during the generation of fix patches, which significantly diminishes the model’s performance.
This gap further underscores the critical role that prompt tokens play in instructing GPT-Neo 1.3B in APR tasks, thereby demonstrating the distinct advantage of prompt tuning over fine-tuning. Compared to GPT-Neo 1.3B, CodeT5+ models are pre-trained on infilling and Seq2Seq generation tasks. As a result, while different templates may influence performance, their impact tends to be minimal.

\vspace{3pt}
\noindent\colorbox{gray!30}{%
\parbox{\dimexpr\linewidth-2\fboxsep\relax}{%
\textbf{Finding 4}: Different templates have a significant impact on prompt tuning, with BP4 performing best for CodeT5+ 770M and BP2 for GPT-Neo 1.3B across most datasets. GPT-Neo 1.3B is more sensitive to these variations, with BP1, BP5, BP6, and BP7 underperforming, while BP2, BP3, and BP4 achieve better results, emphasizing the importance of effective prompt design.
}%
}
\vspace{-4pt}

\begin{table}
  \scriptsize
  \centering
  \caption{Comparison of basic prompt and various knowledge prompts with domain knowledge}
  \label{tab:domain_knowledge}
  \renewcommand{\arraystretch}{1}
  \begin{tabular}{l|l|cc} 
    \toprule 
         \multirow{3}{*}{\textbf{Dataset}} &   \multirow{3}{*}{\textbf{Prompts}} & \multicolumn{2}{c}{\textbf{Model}} \\
    && CodeT5+ & CodeT5+\\
    && 220M & 770M\\
    \midrule 

    \multirow{4}{*}{Defects4J}  &Basic Prompt & 10.94 & 10.94 \\
    &Repair Action & \textbf{14.06} & \textbf{12.50} \\
    &Repair Pattern & 12.50 & 12.50 \\
    &Repair Action + Repair Pattern & 10.94 & 10.94\\
    \midrule 
    \multirow{4}{*}{ManySStuBs4J}  &Basic Prompt  & 55.21 & 56.25\\
    &Bug Type  & 56.25 & 56.25\\
    &AST & \textbf{57.29} & \textbf{57.29}\\ 
    &Bug Type + AST & 56.25 & 56.25\\
    \midrule 
    \multirow{4}{*}{TFix}  &Basic Prompt  & 21.90 & 21.90\\
    &Bug Type  &22.88 & 24.18\\    
    &Error Message  & \textbf{23.53} & 22.55\\
    &Bug Type + Error Message  & 22.55& \textbf{25.82}\\
    \midrule 
    \multirow{4}{*}{xCodeEval}  &Basic Prompt  & \textbf{3.46} & 3.46\\
    &Tags  & 2.96 & 3.70 \\
    &Error Message  & 2.96 & 3.95\\
    &Tags + Error Message  & 2.72& \textbf{3.95}\\
    \bottomrule
  \end{tabular}
  \vspace{-15pt}
\end{table}

\vspace{-6pt}
\subsection{RQ3: Effectiveness of Domain Knowledge}
\label{sec:knowledge} 
In this section, we explore the impact of incorporating code- or bug-related domain knowledge into the knowledge prompts in knowledge prompt tuning. As stated in Section~\ref{sec:kp} and Table~\ref{tab:def}, we design a series of knowledge prompt templates based on different types of domain knowledge and models for different datasets. 
Table~\ref{tab:domain_knowledge} presents the comparison of basic prompts and various knowledge prompts.
In Finding 4, we observe that the positions of [X], [mask], and prompt tokens have a significant impact on the results, particularly for GPT-Neo. To mitigate this effect, in Table~\ref{tab:domain_knowledge}, we
select the basic prompt BP7 (the basic prompt design most similar in structure to the knowledge prompts) as the baseline for comparison with knowledge prompts.
Overall, KPs with various domain knowledge outperform the baseline, indicating the better effectiveness of domain knowledge in prompt tuning than basic prompts. 

\begin{figure}
  \centering 
  \includegraphics[width=0.5\textwidth]{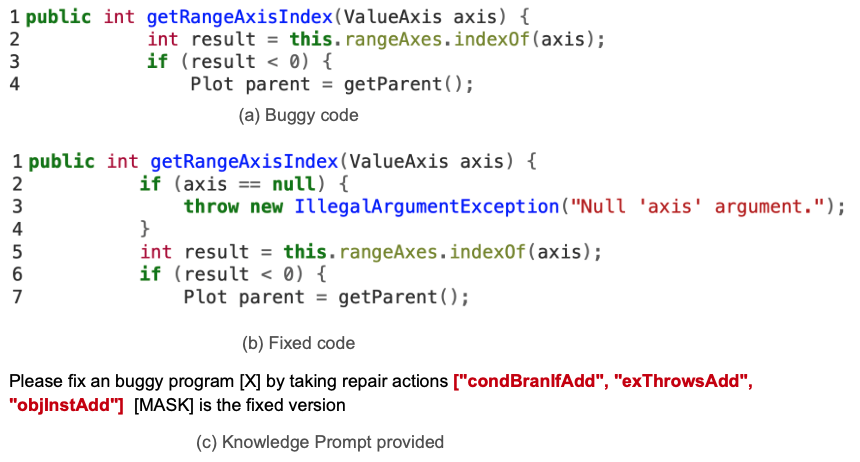} 
  \vspace{-15pt}
  \caption{An example of buggy code from Defect4J successfully fixed by knowledge prompts} 
  \label{fig5} 
  \vspace{-15pt}
\end{figure}

We observe that, with Defects4J, the KP with repair action as the domain knowledge outperforms BP7, showing an improvement of 28.51\% and 14.25\% on CodeT5+ 220M and 770M, respectively. 
An example of buggy code and its fix from Defects4J is shown in Fig.\ref{fig5}. The buggy code in Fig.\ref{fig5}(a) fails to check if the input parameter \textcolor{black}{\texttt{axis}} is null before using it. None of the BPs can repair this example, but the KP, integrated with repair actions, achieves successful repair by guiding the process with CodeT5+ models. Specifically, the repair action \textcolor{red}{\texttt{condBranchAdd}} (conditional branch addition) generates lines 2 and 4 in the fix code (Fig.~\ref{fig5}(b)), while \textcolor{red}{\texttt{exThrowsAdd}} (throw addition) and \textcolor{red}{\texttt{objInstAdd}} (object instantiation addition) collaboratively generate line 3. This example highlights how incorporating repair actions as domain knowledge enables the model to learn richer features in knowledge prompt tuning.

We observe in some cases (e.g. ManySStuBs4J), incorporating a single type of domain knowledge brings substantial improvements. However, when both types are integrated together into one KP, their contributions to prompt tuning are not incremental, and the resulting improvement is less pronounced than when using only one type of knowledge for each model. 
A possible reason is, that when multiple types of domain knowledge are integrated, they might introduce conflicting or noisy information, which can confuse the model. The models might struggle to focus on the most relevant parts, leading to poorer performance than using a single, more coherent source of knowledge.


For xCodeEval, we observe that incorporating tags of potential algorithmic techniques in buggy code and error messages does not lead to significant improvements compared to the baseline. This may be due to irrelevant or noisy domain knowledge since domain knowledge in xCodeEval is generated without being manual validation. When the added knowledge is not closely aligned with the task, it can introduce noise, distracting the model from more relevant features. Therefore, the relevance and quality of domain knowledge are crucial factors to consider when selecting knowledge for prompt tuning.

\vspace{3pt}
\noindent\colorbox{gray!30}{%
\parbox{\dimexpr\linewidth-2\fboxsep\relax}{%
\textbf{Finding 5}: Incorporating code- and bug-related domain knowledge generally improves prompt tuning performance. However, irrelevant or noisy domain knowledge may introduce confusion and reduce the model's effectiveness, highlighting the importance of carefully selecting relevant knowledge.
}%
}
\vspace{-10pt}

\begin{table*}
    \tiny
  \centering
  \caption{Results of prompt tuning and fine-tuning in extreme data scarcity scenarios (C.BLEU represents CodeBLEU)}
  \label{tab:9}
  \renewcommand{\arraystretch}{1}
  \begin{tabular}{c c|c c| cc| c c|c c| cc| c c| cc| cc} 
    \toprule 
     \multirow{3}{*}{\textbf{Language}} & \multirow{3}{*}{\textbf{\textbf{Methods}}} & \multicolumn{8}{c|}{\textbf{CodeT5+ 220M}} & \multicolumn{8}{c}{\textbf{CodeT5+ 770M}} \\
     \cline{3-10}
    \cline{11-18}
      && \multicolumn{2}{c|}{\textbf{1 shot}} & \multicolumn{2}{c|}{\textbf{8 shot}}  & \multicolumn{2}{c|}{\textbf{16 shots}}  & \multicolumn{2}{c|}{\textbf{32 shots}} & \multicolumn{2}{c|}{\textbf{1 shot}} & \multicolumn{2}{c|}{\textbf{8 shot}}  & \multicolumn{2}{c|}{\textbf{16 shots}}  & \multicolumn{2}{c}{\textbf{32 shots}}\\
     && EM & C.BLEU & EM & C.BLEU& EM & C.BLEU & EM & C.BLEU & EM &C.BLEU& EM & C.BLEU & EM &C.BLEU & EM & C.BLEU \\
    \midrule 
     \multirow{2}{*}{Python} & Fine-tune & 0& 12.47 &0&56.64 &0&72.95& 0    &73.83     &0    &21.56  &0& 63.58   & 0.67& 73.76& 0.67&   75.11  \\
    & Prompt tuning                     & 0 &  43.35 &0&66.81&0&75.45 &0.67 &75.84    &0    &57.56  &0.67 &70.24     & 0.67&75.50 & 1.67 & 75.80 \\
    
    \multirow{2}{*}{Java} & Fine-tune & 0 & 6.06 & 70.33&83.54&90.00&87.56  & 174.33 &91.52      & 0  & 18.46  &102.67&84.60    &113.33 & 87.32&203.00 & 92.08\\
    & Prompt tuning                   & 1.67 &30.98 & 75.00&85.21 & 111.67&91.34& 191.00  & 93.25    &3.33&44.28   &104.33&88.43    & 145.00& 91.76& 218.67&  93.24\\
    \multirow{2}{*}{JavaScript} & Fine-tune &0 &9.19 &4.00&47.93      &8.00 &53.28 & 15.00 & 56.62   &0   &15.18   &7.67 & 50.99    &12.33 & 54.66 & 20.00&  56.50\\ 
    & Prompt tuning                        &1.33 & 25.22 & 4.67 &51.01&8.50 &54.82 & 20.33 & 57.47   &1.34& 31.52  &11.33 &52.35     & 19.50& 55.38& 24.67& 57.50 \\
     \multirow{2}{*}{C} & Fine-tune & 0& 15.10&2.00 &70.86 &2.33 & 73.39 & 3.67& 73.96       &0   & 5.72  &6.00& 72.39     &4.00 &73.78 & 4.00& 73.92 \\
    & Prompt tuning                 &0& 61.61 &3.33 &72.81 &3.00 &73.25  & 3.33&73.59      &1.00&61.30  &4.67 &72.43    & 2.33& 49.55 &4.33 & 73.76 \\

    \bottomrule
  \end{tabular}
  \vspace{-5pt}
\end{table*}

\begin{figure*}[t]
  \centering 
  \includegraphics[width=1.0\textwidth]{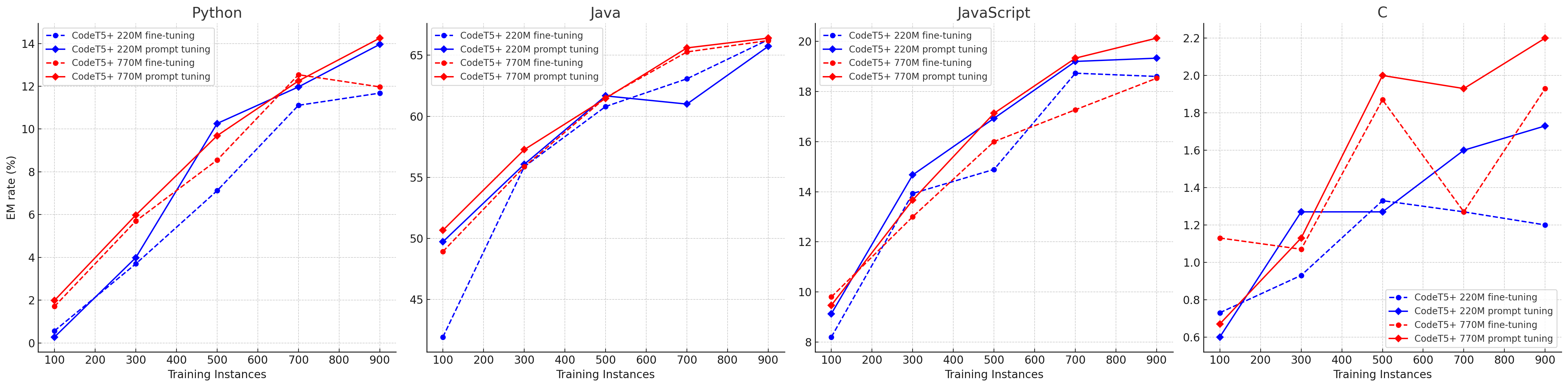} 
  \vspace{-15pt}
  \caption{Results of fine-tuning and prompt tuning across different training set sizes} 
  \label{fig6}
  \vspace{-15pt}
\end{figure*}
\subsection{RQ4: Impact of training dataset sizes}
\vspace{-2pt}
To assess the robustness of prompt tuning across different data sizes, we randomly select subsets of training instances (also known as shots) in quantities of 1, 8, 16, 32, 100, 300, 500, 700, and 900, while the test set is fixed at 500 instances for consistency. To prevent randomization in data selection, we generate each subset three times using distinct seeds, conduct our experiments on each dataset, and present the average results. 
Table~\ref{tab:9} compares prompt tuning and fine-tuning in extreme data scarcity (i.e. with 1,8,14,32 instances), while Fig.~\ref{fig6} shows performance across a range of training set sizes. Overall, the performance of both prompt tuning and fine-tuning tends to improve as the size of the training set increases, despite some fluctuations. Moreover, prompt tuning generally outperforms fine-tuning, especially in scenarios with extremely limited data. 

\subsubsection{Comparison of performance in extreme data scarcity scenarios}
In Table~\ref{tab:9}, prompt tuning generally outperforms fine-tuning across all programming languages in 1, 8, 16, and 32-shot scenarios. 
In the 1-shot scenario, we observe that fine-tuned models exhibit extremely low CodeBLEU scores. In particular, fine-tuned CodeT5+ 220M achieves a CodeBLEU score of only 6.06 on Java, indicating that its generated patches are far from the correct fixes. In contrast, even in the 1-shot scenario, the prompt-tuned CodeT5+ 220M model successfully repairs an average of 1.67 Java test instances. As more data becomes available, prompt tuning maintains a clear advantage.
This indicates that prompt tuning demonstrates greater robustness and effectiveness compared to fine-tuning in scenarios with extreme data scarcity.

\subsubsection{Comparison of performance with increasing training instances}
In this section, we compare the performance trends of prompt tuning and fine-tuning as the number of training instances increases. As shown in Fig.~\ref{fig6}, the EM rates improve for both prompt tuning and fine-tuning across all programming languages as the number of training instances grows.
Moveover, prompt tuning (red plot lines) generally outperforms fine-tuning, achieving up to a 44\% improvement.


The performance trends in C reveal a more fluctuating pattern. Although prompt tuning performs better overall, both methods experience minor declines in performance at certain points. These fluctuations could be due to the inherent complexity of the C dataset, which may require more fine-grained tuning or additional domain knowledge to achieve stable performance. Nevertheless, prompt tuning still demonstrates superior performance across most subsets of training instances.

\vspace{3pt}
\noindent\colorbox{gray!30}{%
\parbox{\dimexpr\linewidth-2\fboxsep\relax}{%
\textbf{Finding 6}: Overall, prompt tuning shows significant improvement over fine-tuning across various training set sizes, particularly in extreme data scarcity scenarios. Both methods generally improve as the training set size increases. 
}%
}
\vspace{-8pt}

%% file: disscussion.tex
\vspace{-2pt}
\subsection{Implications}

\subsubsection{Soft and hard Prompts}
Our experiments indicate that soft prompts generally perform better than hard prompts in APR tasks. As mentioned in our study, hard prompts often need to be manually defined by human experts, but these designed prompts can be suboptimal. Soft prompts address this issue by tuning the prompt values during training. However, this also makes soft prompts less interpretable and requires more data to achieve optimal performance, especially when the soft tokens are initialized randomly. Therefore, researchers should make the decision between soft prompt and hard prompt based on specific needs and data availability.

\subsubsection{Initialization methods of soft prompts}
Our experiments indicate that while the performance of SBP$_{\text{initialized}}$ and SBP$_{\text{random}}$ is typically similar, one can exhibit an advantage over the other depending on the different model and dataset. This is because vocabulary-based initialization in SBP$_{\text{initialized}}$ benefits from vocabulary-based initialization, providing a stable starting point that requires less training time and data. In contrast, SBP$_{\text{random}}$, with randomly initialized tokens, offers greater flexibility to explore alternative local optima, though it generally requires more data and training time. Therefore, for researchers the choice between SBP$_{\text{initialized}}$ and SBP$_{\text{random}}$ should account for factors like dataset size, task complexity, and available computational resources.

\subsubsection{Prompt template design}
Our experiments show that the design of prompt templates significantly affects the performance of GPT-Neo 1.3B. As the sequence of the input buggy program, prompt tokens, and mask tokens is critical for guiding the model effectively, especially for the models that are not pre-trained on specific tasks, optimizing the prompt template through experimentation is essential. Researchers should design and evaluate prompt templates based on the specific type of models and datasets in their work.

\subsubsection{Incorporating Domain Knowledge}
Our experiments reveal that incorporating code-or bug-related domain knowledge specific to individual buggy code generally enhances prompt tuning performance by helping the model better understand the context of each bug. However, irrelevant or noisy information can reduce effectiveness, highlighting the importance of carefully selecting and validating the domain knowledge before integration. Researchers should prioritize domain knowledge that directly relates to the specific characteristics of the buggy code and avoid using irrelevant domain knowledge. 
\vspace{-4pt}
\subsection{Threats To Validity}

\subsubsection{Construct Validity}
Using Exact Match as our primary metric may not fully capture the actual fix ratio. To address this, we also use syntactically correct patches and CodeBLEU as additional metrics. Furthermore, our prompt template design may not be optimal for the experiments. To mitigate this, we create 7 basic prompt templates and 12 knowledge-based templates, evaluating prompt-tuning performance across these different designs.

\subsubsection{Internal Validity}
Each experiment run might yield slightly different results due to inherent randomness in the training process. Since we focus on data scarcity scenarios, we sample 1\% of the data from datasets with large sizes, but this randomization of sampling may affect the results of evaluation. To mitigate this issue, for each prompt template design, model, and dataset, we run our experiments on fine-tuning and prompt-tuning three times and take the average as the final result to mitigate the impact of randomization in the model’s internal processes and data sampling. 

\subsubsection{External Validity}
Our experimental results may be applicable to specific datasets and models only. 
We mitigate this threat by selecting six diverse datasets, including four different languages, and three models of varying sizes and types. 
However, we may need more evaluation in the future to conclude that our results could be generalizable to all datasets and models.

\vspace{-4pt}

%% file: relatedwork.tex
\subsection{Prompt Tuning}

Prompt tuning is an advanced method that trains a pre-trained model using a small amount of task-specific data along with a set of prompts. 
Prompt tuning is first to be applied to complete NLP tasks and it shows efficiency since it aligns the objectives of downstream tasks and pre-tuning. B. Lester and colleagues~\cite{prompttuning} demonstrated that through prompt-tuning, T5 could achieve comparable quality as fine-tuning as the model size increases, and their approach to prompt-tuning T5 surpasses few-shot prompt design using GPT-3. 
Recently, C. Wang and colleagues \cite{no} show that prompt tuning enhances CodeBERT and CodeT5's performance in code intelligence tasks (i.e., defect detection, code summarization, and code translation), surpassing fine-tuning especially in low-resource settings, suggesting its viability as a superior tuning method for scarce data scenarios. However, the effectiveness of prompt tuning in APR is underexplored. In our work, we aim to close this gap.
\vspace{-4pt}
\subsection{Leverage of domain knowledge}
Prior works have shown that incorporating domain knowledge into model design significantly boosts performance, particularly in data scarcity scenarios. In NLP, KnowPrompt~\cite{knowprompt} and Knowledgeable Prompt-tuning (KPT)~\cite{knowledgeable} demonstrate how leveraging domain-specific knowledge in prompts enhances relation extraction and text classification accuracy, especially in low-resource settings.
For code intelligence tasks, integrating domain knowledge in pre-training, prompt engineering, and fine-tuning has proven effective. For instance, UnixCoder~\cite{guo-etal-2022-unixcoder} outperforms prior models by incorporating ASTs to enrich code representation in the pre-training stage. In APR, RAP-Gen~\cite{rapgen} uses retrieval-augmented prompts for patch generation, and TFix~\cite{tfix} combines error type and message with buggy code to fine-tune the T5 model~\cite{t5}.
Our work differs in that we incorporate both code- and bug-related knowledge into prompt tuning for APR and evaluating multiple types of domain knowledge.

%% file: conclusion.tex
This work demonstrates that prompt-tuning generally outperforms fine-tuning in APR under data-scarcity scenarios, particularly for models not pre-trained on specific tasks. Incorporating code- or bug-related domain knowledge (e.g., repair actions,  ASTs of buggy node, and bug types) into prompts further enhances prompt-tuning performance.
Through a comprehensive comparison of various prompt types and template designs, we provide insights and practical suggestions for researchers in future studies.
Although our evaluation is only done with four datasets, three relatively small pre-trained code models, and limited prompt templates, we highlight the significant potential of adapting prompt tuning for APR tasks over fine-tuning, especially in data scarcity scenarios. Furthermore, we offer insights for future advancements in integrating code- or bug-related domain knowledge to further enhance APR performance.
Our source code and experimental data are publicly available at: \url{https://github.com/Cxm211/k-prompt}

%% file: acks.tex
This research is supported by the Ministry of Education, Singapore under its Academic Research Fund Tier 3 (Award ID: MOET32020- 0004). Any opinions, findings and conclusions or recommendations expressed in this material are those of the author(s) and do not reflect the views of the Ministry of Education, Singapore.